\documentclass[journal=ancac3,manuscript=article]{achemso}
\usepackage{amsmath}
\usepackage{graphicx}
\usepackage{xcolor}

\title{Surface Rearrangement and Evaporation Kinetics of Supported Gold Nanoparticle Catalysts}
\author{James P. Horwath}
\affiliation{University of Pennsylvania, Department of Materials Science and Engineering, Philadelphia, PA 19104}
\author{Colin Lehman-Chong}
\affiliation{University of Pennsylvania, Department of Chemical and Biomolecular Engineering, Philadelphia, PA 19104}
\author{Aleksandra Vojvodic}
\affiliation{University of Pennsylvania, Department of Chemical and Biomolecular Engineering, Philadelphia, PA 19104}
\email{alevoj@seas.upenn.edu}
\author{Eric A. Stach}
\affiliation{University of Pennsylvania, Department of Materials Science and Engineering, Philadelphia, PA 19104}
\email{stach@seas.upenn.edu}

\begin{document}

\begin{abstract}
    Heterogeneous catalysts consisting of supported metallic nanoparticles typically derive exceptional catalytic activity from their large proportion of under-coordinated surface sites which promote adsorption of reactant  molecules. Simultaneously, these high energy surface configurations are unstable, leading to nanoparticle growth or degradation, and eventually a loss of catalytic activity. Surface morphology of catalytic nanoparticles is paramount to catalytic activity, selectivity, as well as degradation rates, however, it is well-known that harsh reaction conditions can cause the surface structure to change. Still, limited research has focused on understanding the link between nanoparticle surface facets and degradation rates or mechanisms. Here, we study a model Au supported catalyst system over a range of temperatures using a combination of \textit{in situ} Transmission Electron Microscopy, kinetic Monte Carlo simulations, and density functional theory calculations to establish an atomistic picture of how variations in surface structures and atomic coordination environments lead to shifting evolution mechanisms as a function of temperature. By combining experimental results, which yield direct observation of dynamic shape changes and particle evaporation rates, with computational techniques, which enable understanding the fundamental thermodynamics and kinetics of nanoparticle evolution, we illustrate a two-step evolution mechanism in which mobile adatoms form through desorption from low-coordination facets and subsequently evaporate off the particle surface. By understanding the role of temperature in the competition between surface diffusion and evaporation, we are able to show how individual atomic movements lead to particle-scale morphological changes, and rationalize why evaporation rates vary between particles in a system of nearly identical nanoparticles.
\end{abstract}

\maketitle

Supported metal nanoparticles (NPs) are commonly used in industrial catalysis.  As heterogeneous catalysts, under-coordinated active sites on the surfaces of  nanoparticles promote adsorption of reaction gas molecules and facilitate chemical reactions to enhance both catalytic activity and selectivity.
However, the high surface free energy associated with these under-coordinated surfaces and instability of under-coordinated active sites simultaneously drive nanoparticle evolution to reduce the total free energy of the system by reducing the catalyst surface area. Maintaining catalytic activity for important commercial reactions therefore commonly relies on preventing this coarsening, and keeping nanoparticles small\cite{liu_metal_2018}.  Research focusing on coarsening and degradation mechanisms in supported nanoparticles has focused on the use of mean-field theories for decades.  While this approach is helpful for developing an analytical, intuitive model which can be compared against simple experiments, several key assumptions limit the use of these models for specific applications and real systems with large complexity.  Mean-field theories which describe diffusional interactions between particles and their surrounding field rely on the assumption of infinite separation between particles to avoid accounting for direct interactions between particles, and disregard any surface energy anisotropy by studying the behavior of a smooth, (truncated) spherical particle.\cite{balluffi_kinetics_2005,voorhees_theory_1985}.  However, these assumptions are not valid in real systems with complex, heterogeneous and anisotropic particles which are the key features of these systems responsible for their unique chemical and catalytic behaviour. 

In reality, it is well known that nanoscale particles are commonly faceted and that the surface energy of each facet is not the same and can change with environmental conditions \textit{e.g.} temperature, oxidizing or reducing conditions, reacting gases\cite{williams_surface_1989}.  Therefore, understanding the competition between degradation mechanisms in model catalysts requires moving beyond mean-field theories and focusing on local changes which cause deviations from the expected mean-field behavior.
In our previous work, Au NPs heated to 900 $^\circ$C were found to evolve by a competition between evaporation out of the system and adatom diffusion along the substrate\cite{horwath_quantifying_2021}.  While high temperatures are required to increase the vapor pressure of solid gold such that evaporation is favorable, we expect that diffusion of adatoms along the particle surface is still possible at lower temperatures, and that this diffusion could lead to important morphological changes.  Several recent studies have focused on using atomic-resolution scanning transmission electron microscopy (STEM) to quantify the morphology of faceted nanoparticles. Using these  methods, researchers have observed beam-induced adatom diffusion on faceted surfaces, motion of entire layers leading to stochastic rearrangement of particle morphology, and the influence of edge atoms on the concentration of diffusing atoms\cite{de_wael_measuring_2020,liu_three-dimensional_2021}.  Others have used time-resolved STEM to observe facet changes as a nanoparticle is subjected to oxidizing and reducing environments and showed reaction-induced surface roughening\cite{altantzis_three-dimensional_2019}.  Catalysis studies have also shown enhanced reaction rates on specific crystal surfaces and surface-dependent coarsening rates in supported nanoparticles, further demonstrating the importance of understanding how catalyst performance and degradation rates depend on nanoscale morphology\cite{yuan_direct_2018-1,bi_facet_2011,chmielewski_reshaping_2019}.

In this manuscript, the goal is to understand how the dominant evolution mechanisms in gold nanoparticles change as a function of temperature, and to elucidate the role of surface structures on evolution kinetics.  We use time-resolved \textit{in situ} transmission electron microscopy (TEM) to understand the mechanisms which contribute to nanoparticle degradation as a function of temperature.  Motivated by our experimental findings, which suggest temperatures above 750 $^\circ$C are required to activate the evaporation process on Au NP, we performed kinetic Monte Carlo (KMC) simulations to achieve atomic scale understanding of the mechanisms which lead to particle-scale shape changes.  By investigating how fundamental particle properties, \textit{e.g.} the coordination number of atoms on different surfaces, lead to stochastic evolution processes we are able to clearly observe the growth of new surface layers, the formation of non-equilibrium facet orientations, and the relationship between the particle structure and the evaporation rate.  Further, density functional theory (DFT) calculations are applied to study the low energy reaction pathways which lead to nanoparticle degradation.  This combination of time-resolved experimental data, atomistic kinetic simulations, and first principles thermodynamic calculations allow us to build a comprehensive picture of the relationship between particle surface properties, environmental conditions, and evaporation rates.

\section{Results and Discussion}

\subsection{Temperature-dependent Evolution Behavior}
\begin{figure}[h]
    \centering
    \includegraphics[width=\textwidth]{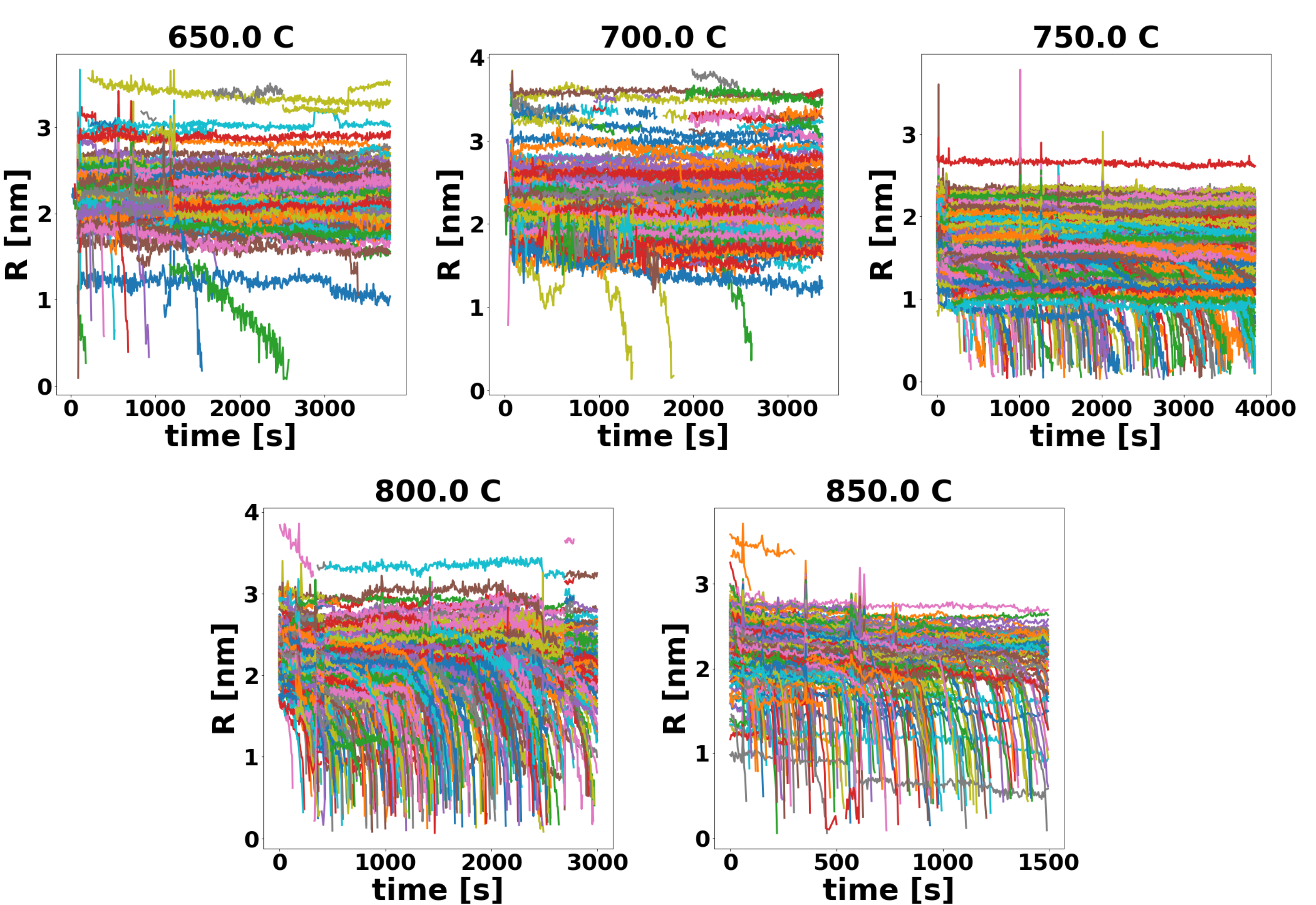}
    \caption{Size (R) \textit{vs.} time profiles for all particles tracked in experimental TEM image series.  Each plot shows the real time evolution of nanoparticles as a function of time and temperature.}
    \label{fig:growt_v_t}
\end{figure}

Our model catalyst consisted of 5~nm Au nanoparticles supported on a silicon nitride \textit{in situ} heating chip.  The system was heated to high temperature within the TEM, and images were collected as a function of time (see Methods section for further details).  Nanoparticle sizes were measured from TEM images, and particle size as a function of time was tracked for each particle.
Experimental data show that below 750 $^\circ$C particle size changes very little throughout the experiment.  At 600 – 700 $^\circ$C a few nanoparticles evaporate and there are some examples of particle coalescence (seen as discontinuous growth/disappearance), however, most particles do not change size significantly (Figure~\ref{fig:growt_v_t}).  At 750 $^\circ$C many of the particles in view begin to rapidly evaporate, and the process accelerates as the temperature continues to increase.  At temperatures of 750 $^\circ$C and above there is no clear evidence for particle coalescence, and we find that evaporation dominates.

\begin{figure}[h]
    \centering
    \includegraphics[width=\textwidth]{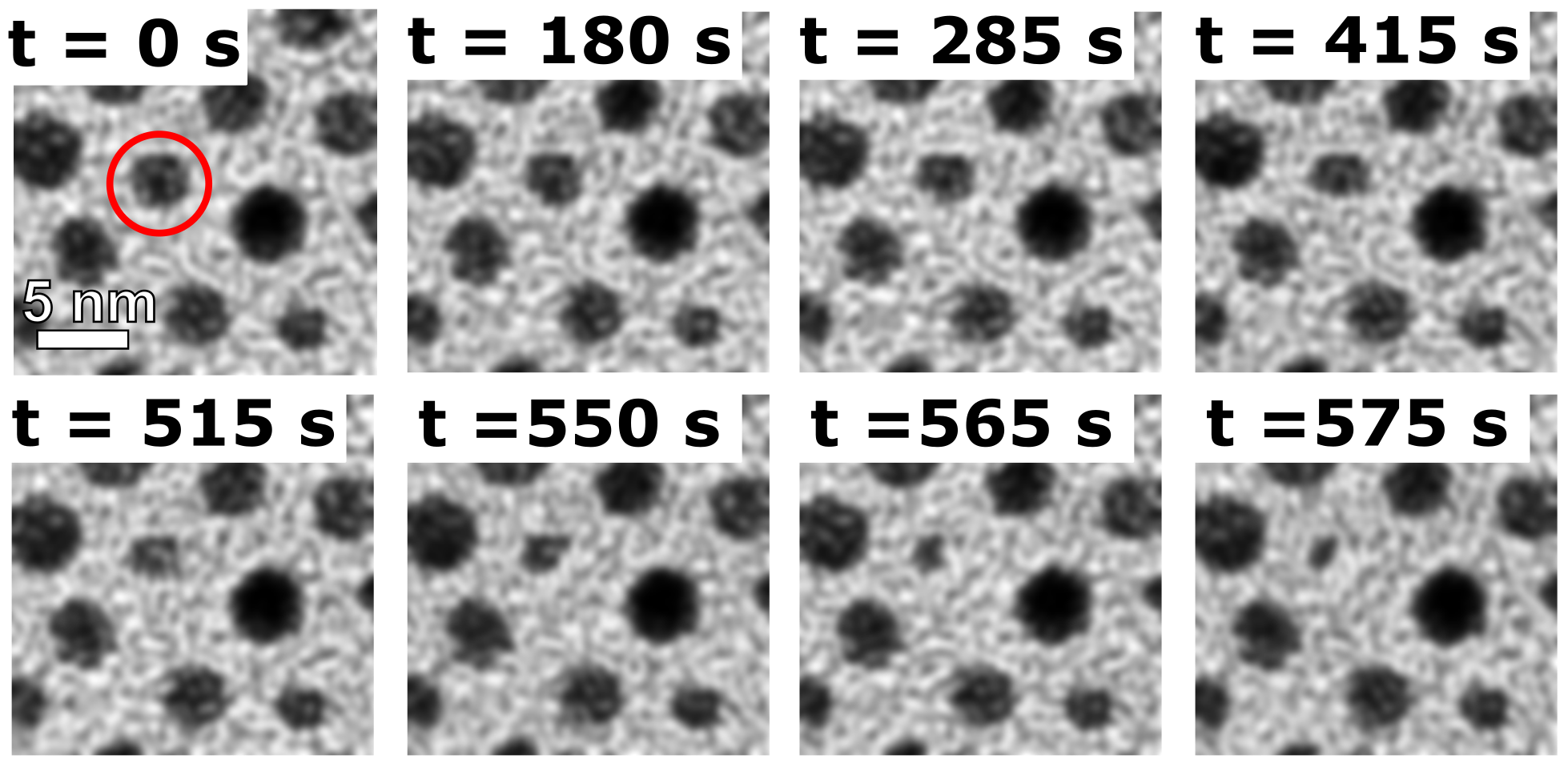}
    \caption{Evolution of an evaporating particle (circled in red at t = 0 s) through time.  Faceted notches form at the surface of the particle and trigger fast evaporation. Scale bar represents 5 nm.}
    \label{fig:evap_images}
\end{figure}

By directly observing changes to the shapes and sizes of individual nanoparticles, we can identify patterns in the evolution behaviors as nanoparticles shrink. Figure~\ref{fig:evap_images} shows the evolution of an evaporating nanoparticle which clearly suggests preferential evaporation from specific crystallographic facets, as throughout the evaporation processes flat facets and sharp angular notches form at the particle surface.

\begin{figure}[h]
    \centering
    \includegraphics[width=\textwidth]{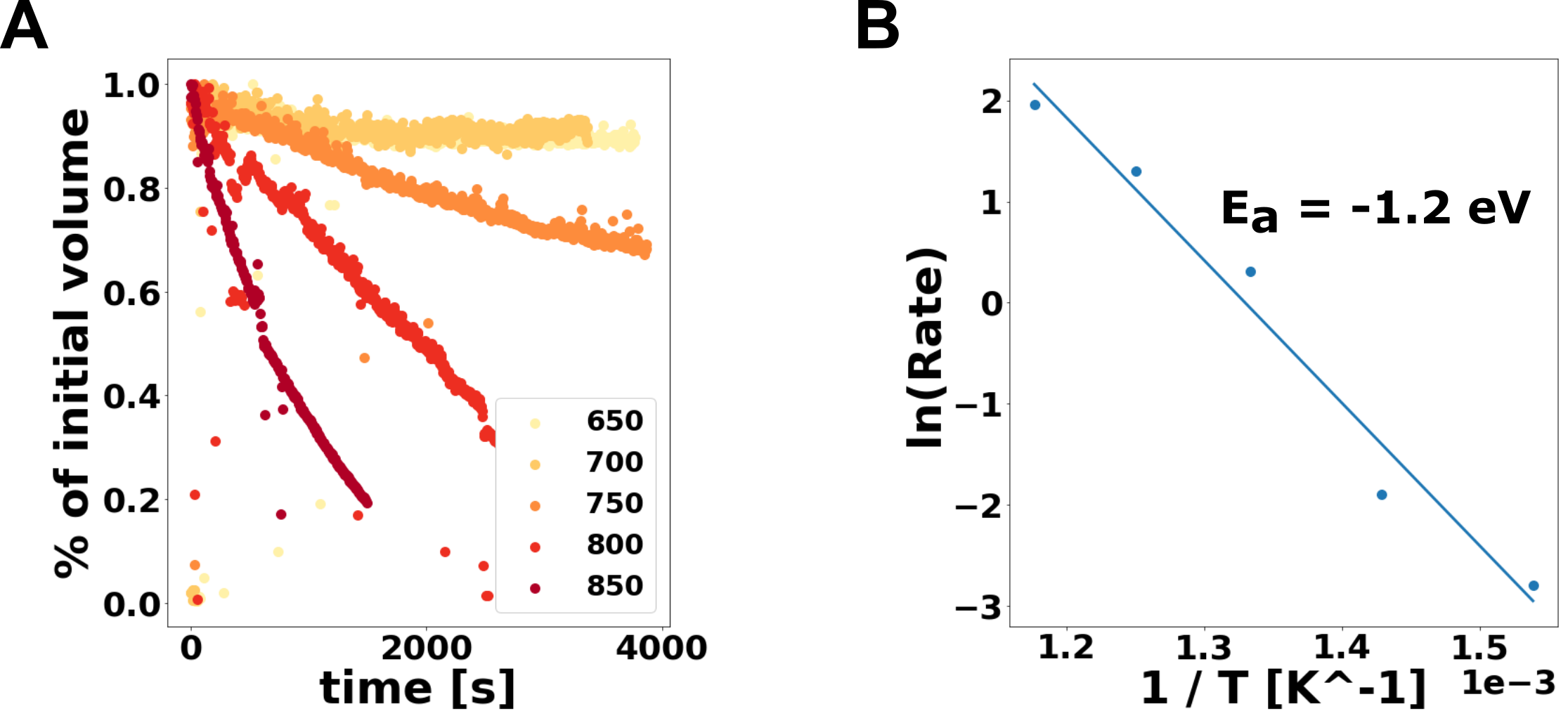}
    \caption{A.) The evaporation rate is determined by fitting the slope of the the volume of Au in the system as a function of time. Very little mass is lost below 700~$^\circ$C. B.) The evaporation rates fit and Arrhenius form, such that an activation barrier for evaporation can be extracted.}
    \label{fig:activation}
\end{figure}

To better understand the kinetics and temperature-dependence of the system, the evaporation rate was plotted as a function of temperature.  Assuming that evaporation is the only process which removes Au mass from the model catalyst, the total volume of gold in the system (assuming spherical particles) was calculated in each frame of the TEM dataset, fit to a line as a function of time, and the evaporation rate was taken as the slope of the line (Figure~\ref{fig:activation} A).  For all temperatures, a linear relationship between total Au volume as a function of time showed clear agreement with experimental data.  A dramatic increase in evaporation rate is observed as the temperature approaches and exceeds 750~$^\circ$C (Figure~\ref{fig:activation}A). We find an agreement with the familiar Arrhenius form (Figure~\ref{fig:activation}B) suggesting that the evaporation of Au from nanoparticle surfaces is kinetically activated and not a dominant reaction at all temperatures.  The extracted activation energy for evaporation of Au was found to be 1.2~eV. We performed DFT calculations showing that this activation energy value is on the same order as, but still significantly less than, the calculated enthalpy of sublimation from any low energy Au surface (see table of desorption energies in Supplemental Table 1 and Supplemental Figure 1).  This combination of experimental observations and first-principles calculations demonstrate that a more complicated mechanism than simple desorption is responsible for nanoparticle evolution.

\subsection{Atomic-scale Morphological Changes}
\begin{figure}[h]
    \centering
    \includegraphics[width=\textwidth]{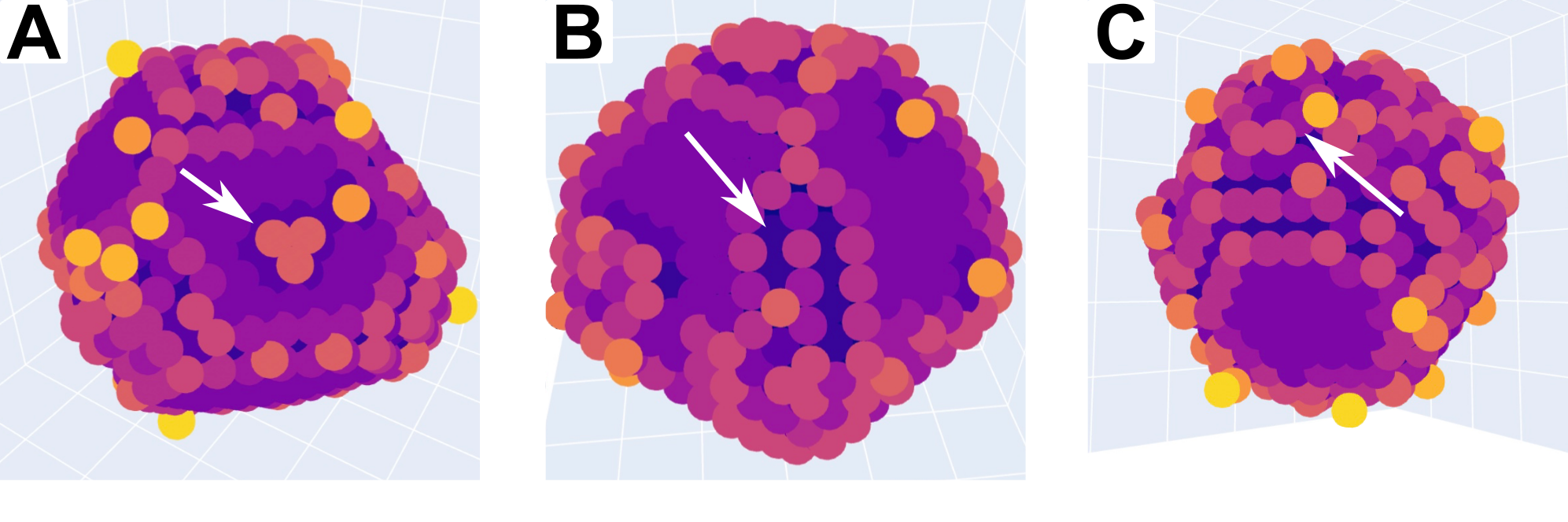}
    \caption{KMC results show the evolution of an evaporating nanoparticle.  In A.), a mobile cluster of adatoms is shown on a \{1 1 1\} surface. B.) shows the development of a \{1 1 0\} facet through the removal of edge atoms.  C.) shows the growth of a new layer on top of the original particle surface.}
    \label{fig:kmc}
\end{figure}

Motivated by our experimental observations of evaporation as an activated process and clear transitions between stable surface structures in coalescing and evaporating nanoparticles, we carried out kinetic Monte Carlo (KMC) simulations to stochastically model the atomic processes that lead to morphological changes in the nanoparticles.  Our simulation starts with model particle corresponding to an equilibrium-shape Au nanoparticle with atoms fixed on a perfect face-centered cubic (FCC) lattice and considers diffusion of surface atoms to unoccupied neighboring lattice sites (diffusion into vacuum, or lattice sites with coordination number zero, represent evaporation).  The nanoparticle is modeled as an without considering the support even though the experimental system consists of supported particles.  Our previous research demonstrated that Au nanoparticle evolution (on an identical support) is dominated by evaporation from free surfaces with only a minimal contribution from diffusion along the support surface, and, due to our particle deposition method, the particle-support contact area is very small\cite{horwath_quantifying_2021}.  Directly modeling metal-support interactions would be more important in systems where NP wet the support - here, the morphology of the entire particle could change as the contact area changes.  Activation energies are based on a simple broken bond model described in the Methods section.
Figure~\ref{fig:kmc} shows the model particle at several points in the KMC simulation.  Using three dimensional animations to track step-by-step changes in the structure, we identified three unique processes by which the particle morphology changes: the formation of adatom clusters on large facets (Figure~\ref{fig:kmc}A), the appearance of facets not present in the initial structure (Figure~\ref{fig:kmc}B), and cluster pinning at defect sites which leads to the growth of new layers (Figure~\ref{fig:kmc}C).  Due to the relatively large number of broken bonds required for evaporation compared to diffusion and the exponential decrease in event probability as a function of coordination number, most KMC events at all temperatures correspond to adatom diffusion on existing crystal surfaces.  We find that when three or more adatoms meet on a surface they tend to form a mobile cluster that migrates on large (1 1 1) surfaces.  In some cases, these clusters are transient and only exist for short time periods, but in others they stay together while their configuration rapidly fluctuates.  As more adatoms join the cluster, it becomes more stable in shape and less mobile on the surface. This is because cluster motion would require concerted movements of many atoms which is unlikely in our simple model. However, we do see that small clusters become pinned when they encounter facet edges or surface vacancies.  The pinning of these clusters at high energy defect sites allows them to remain stationary and continue to grow into a low energy configuration through the addition of more adatoms, enabling the growth of new layers/terraces on top of the initial structure.  In agreement with the well-known Burton-Cabrera-Frank model for step-flow kinetics, we see that under-coordinated edge and corner atoms are removed first, on average, which triggers removal of the remaining edge atoms\cite{burton_growth_1951-1,burton_crystal_1949}.  The collective removal of the edges between two facets forms new facets with different crystallographic orientation.  For example, removal of the edge between \{1 1 1\} and \{1 0 0\} facets creates a \{3 1 1\} plane, while the removal of an edge between two \{1 1 1\} facets creates a \{1 1 0\} plane.

In order to more clearly understand the changes induced by the formation of new layers and facets, and to understand how this relates to the behavior of evaporating nanoparticles we carried out diffraction simulations using the XrDebye module of the ASE python library.  The Debye equation calculates diffraction patterns based on the wave interference caused by all unique pairs of atoms in the system; summing over all interference patterns builds a diffraction pattern which agrees with experimental results and calculations based on other methods, and does not require a mathematical description for shape and structure factors\cite{derlet_calculation_2005}.  Simulated diffraction patterns are displayed as a function of time in Supplemental Figure 2A. Using the assumption that, for a monometallic particle, changes in diffraction peak intensities corresponds to changes in the volume fraction of diffracting surfaces, peak intensity ratios with respect to the dominant \{1 1 1\} diffraction intensity as a function of time were used to qualitatively describe morphological changes in the nanoparticle during evaporation\cite{zhang_identification_2020}.  Our results show increases in the relative intensities of both (2 0 0) and (3 1 1) peaks. While the (2 0 0) intensity increases linearly through the course of the simulation, the (3 1 1) peak increases linearly with a shallow slope initially, before rapid increase at later times.  The rapid increase in (3 1 1) intensity coincides very closely with the time step at which the particle evaporation rate begins to rapidly increase (Supplemental Figure 2B and C) suggesting that the formation, or exposure, of \{3 1 1\} is correlated with an increased evaporation rate.

\begin{figure}[h]
    \centering
    \includegraphics[width=\textwidth]{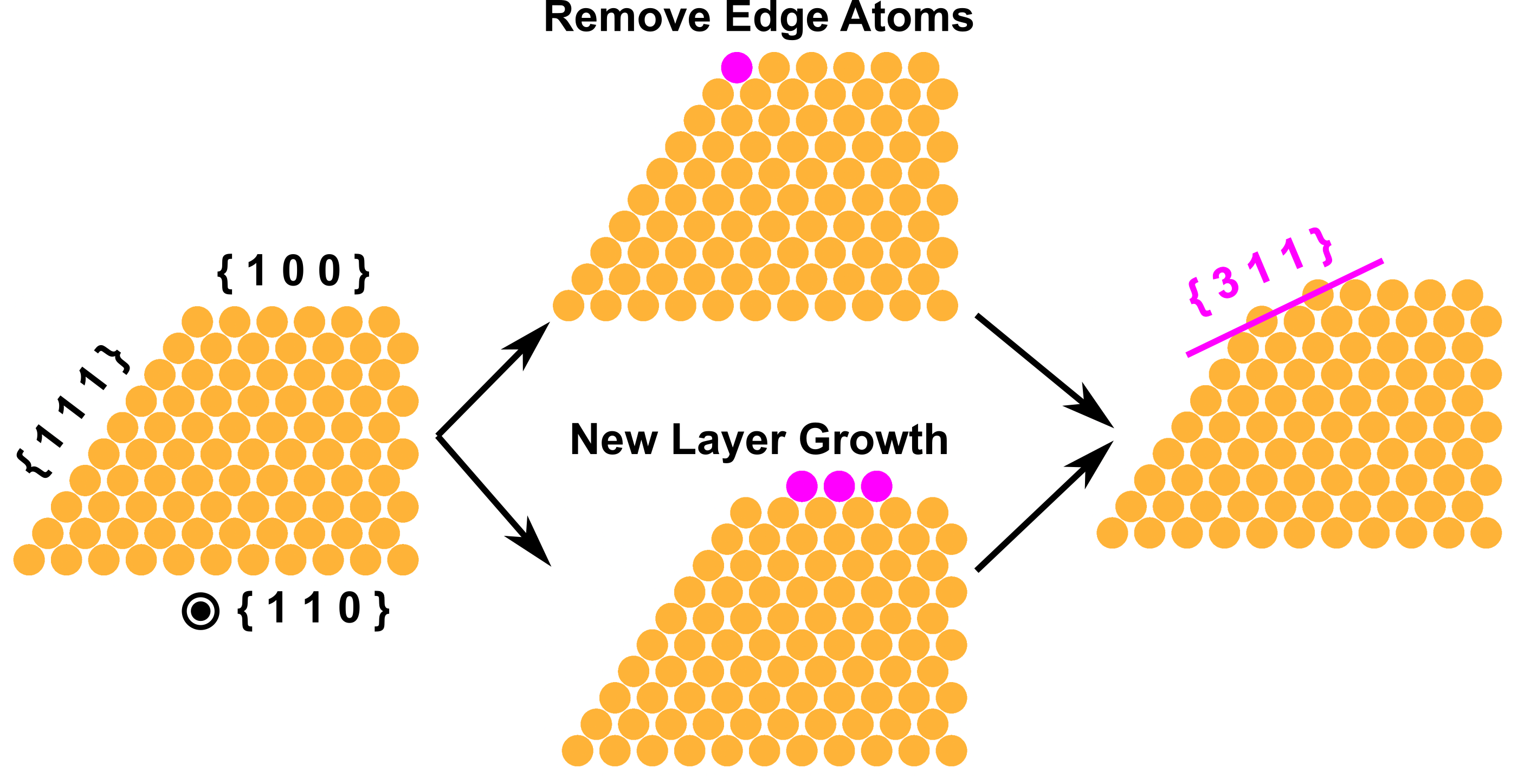}
    \caption{Schematic view of how \{$n$ 1 1\} facets can be formed either by removing atoms from the particle edge, or the growth of new layers.  The purple atoms represent the atoms which are added/removed to make the purple \{3 1 1\} plane in the right-most structure.}
    \label{fig:build_311}
\end{figure}

The \{$n$ 1 1\} surfaces (for odd $n$) are vicinal surfaces for the \{1 0 0\} family.  This means that they are slightly misoriented from the low index \{1 0 0\} planes, and that increasing $n$ corresponds to increasing the width of \{1 0 0\} terraces.  The surface science literature shows both experimental and theoretical verification for the formation of specific, stable vicinal surfaces during the roughening process to help stabilize high energy or highly misoriented surfaces. For example, Bartolini demonstrated that \{3 1 1\}, \{5 1 1\} and \{11 1 1\} are “magic” vicinals for FCC crystals which are exceptionally stable in evolving surfaces\cite{bartolini_magic_1989}.  It is suggested that these orientations are most stable because they are able to accommodate small local distortions which produce a “corrugated surface”, rather than a surface with many high energy step edges.  Our KMC model shows two mechanisms by which \{$n$ 1 1\} planes can develop during the diffusion/evaporation evolution process.   First, the removal of atoms at the edges between \{1 0 0\} and \{1 1 1\} faces forms \{3 1 1\} planes directly (Figure~\ref{fig:build_311}).  Further removal of edge atoms would increase $n$ and increase the width of \{1 0 0\} terraces.  Secondly, the pinning of mobile adatom clusters at edge defects on the particle surface, and subsequent growth to form a new layer, can form this type of surface.  In agreement with this mechanistic picture of \{3 1 1\} formation via fast evaporation from the removal of edges, formation of steps, and subsequent layer evaporation, the angle between the evaporating surfaces in our experimental images of evaporation corresponds very closely to the angle between \{1 0 0\} and\{1 1 1\} faces (Figure~\ref{fig:evap_images}, t = 180 s), where edge removal would form \{$n$ 1 1\} vicinal surfaces.  Moreover, the side length of the notch which initially forms corresponds with the lattice parameter of the Au FCC structure suggesting a step-wise evolution pattern.

\subsection{Desorption Mechanism and Kinetic Equilibrium}
\begin{figure}[h]
    \centering
    \includegraphics[width=\textwidth]{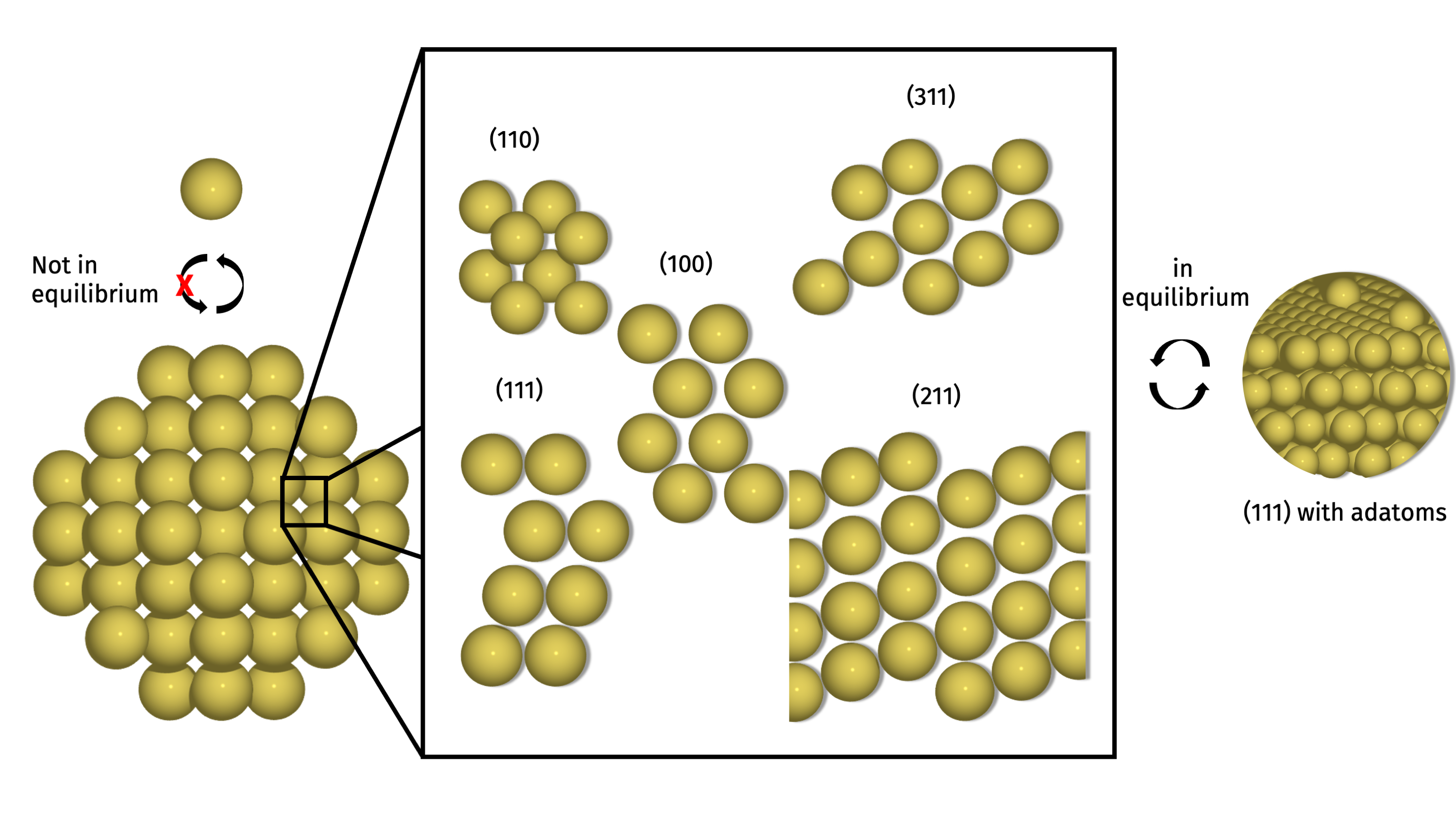}
    \caption{Schematic illustrating the evaporation mechanism whereby surface facets are in equilibrium with (111) adatoms that undergo rate-limiting desorption into the vacuum where they are swept away before equilibrium can be established.}
    \label{fig:schematic_1}
\end{figure}

DFT calculations were utilized to provide an atomistic explanation for nanoparticle evaporation via preferential desorption from specific crystal facets. The model consisted of calculated surface defect formation energies ($\Delta G_{defect}$) and gold adatom adsorption energies ($\Delta G_{adsorption}$).  We considered low Miller index (1 0 0), (1 1 0), (1 1 1), (2 1 1) and (3 1 1) facets which would be expected to dominate nanoparticles in the size and temperature ranges considered experimentally.  Since initial surface defect formation energies and adsorption energies of adatoms on the (1 1 1) surface sites were significantly higher than the experimental activation barrier (Supplemental Table 1), we hypothesized that the evaporation process under these conditions is driven by an adatom formation step followed by a rate-limiting adatom desorption step and that these processes are entropically driven at elevated temperatures.  Guided by the experimental system design in which nanoparticles are under continuously pumped at high vacuum conditions leading to non-conservative conditions, the DFT model suggests an equilibrium between the existence of surface facets and mobile (1 1 1) adatoms (Figure~\ref{fig:schematic_1}) and further assumes that evaporated gold atoms cannot redeposit on the surfaces of the nanoparticles. 

\begin{figure}[h]
    \centering
    \includegraphics[width=\textwidth]{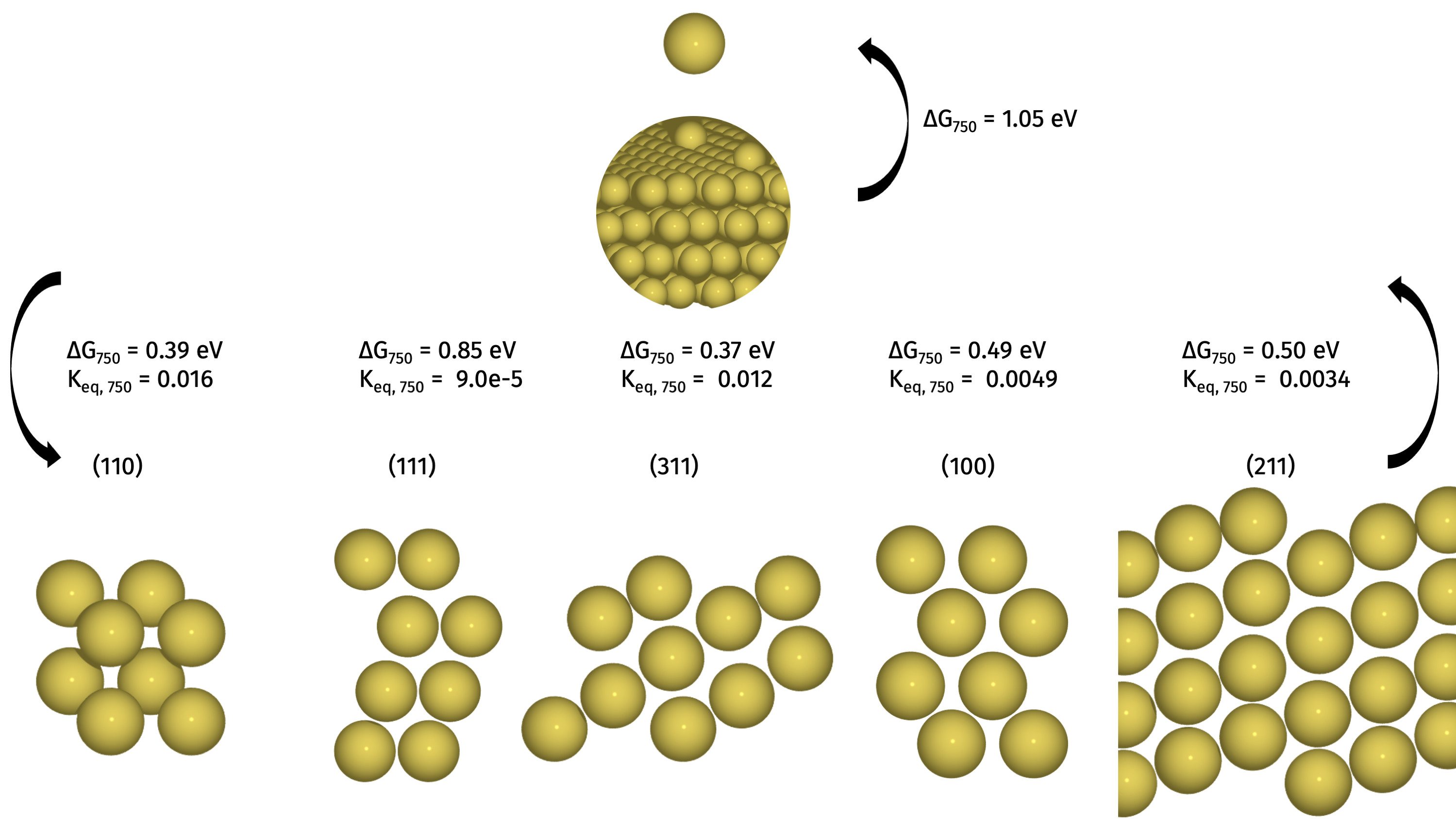}
    \caption{Free energy changes and equilibrium constants associated with the formation of (111) adatoms from the surface atoms of the (111), (110), (100), (211), and (311) surface facets. At the top is also the calculated free energy change for the desorption of a (111) adatom into the gas phase at 750 C.}
    \label{fig:evap_eq_1}
\end{figure}

Hence, our model of the evaporation process is as follows: upon heating, the nanoparticle reaches a point at which the equilibrium constant for the adatom forming reaction step becomes sizeable enough for substantial adatom formation and corresponding surface atom removal.  These adatoms form from the low-coordination surfaces (steps, kinks, \textit{etc.}) as evidenced by the relatively low formation energies for the (3 1 1) and (1 1 0) surfaces shown in Figure~\ref{fig:evap_eq_1}.  This process is ultimately entropically driven; the adatoms and gas phase gold atoms have significant translational entropy contributions and at elevated temperatures they become more stable relative to the higher index surface facets.  Formation energies in Figure~\ref{fig:evap_eq_1} are calculated using $\Delta G_{750}= \Delta G_{defect}+\Delta G_{adsorption}$, where $\Delta G_{750}$ is the energy required to take a gold atom from its lattice position in the surface and place it as an adatom adsorbed on the (1 1 1) surface at 750~$^\circ$C. $\Delta G_{defect}$ is the free energy change associated with creating a defect in the surface by removing an atom to the gas phase, and $\Delta G_{adsorption}$ is the adsorption energy of a gold adatom on the (1 1 1) surface relative to the gas phase. Equilibrium constants are calculated as $K_{750}=e^{\frac{-\Delta G_{750}}{k_{B}T}}$. 
Calculated adsorption energies for gold adatoms on the three-fold and bridge sites of the (1 1 1) surface confirm that the activation barrier for the surface diffusion of these adatoms is very small (0.01~eV), implying that they have two essentially unhindered translational degrees of freedom and substantial entropy at the temperatures relevant to the experiments (Supplemental Table~2).  The adatoms then can undergo desorption with $\Delta G \approx 1$~eV at 750~$^\circ$C, in direct agreement with the experimentally determined activation barrier, supporting this two-step picture of adatom formation and desorption from \{1 1 1\} terraces (Figure~\ref{fig:evap_eq_1}). To confirm the DFT findings, we calculated the equilibrium concentration of \{1 1 1\} adatoms on model nanoparticle surfaces directly from the KMC simulations described above.  Here, \{1 1 1\} adatoms are defined as atoms in a KMC step with a coordination number of three.  Supplemental Figure~3 shows a plot of the adatom concentration (calculated as the number of \{1 1 1\} adatoms divided by the total number of surface atoms), where we can clearly see that the equilibrium KMC adatom concentration at all temperatures converges to 1$\%$ in agreement with equilibrium constant $K_{eq,750}=0.012$.

For temperatures of 750~$^\circ$C and below, where the evaporation rate is low and surface roughening occurs slowly, we see experimental evidence for preferential evaporation from specific crystallographic surfaces.  Our experimental data show the evolution of an evaporating particle (Figure~\ref{fig:evap_images}).  The evaporation process begins by the formation of a notch in the edge of the nanoparticle at a nearly 90$^\circ$ angle.  The notch continues to grow and forms a step edge while the surrounding particle surface becomes smoother.  Subsequently, the step edge rapidly disappears, and we see the formation of a flat particle surface.  The process continues in the same way until evaporation is complete.  It is clear that certain crystallographic faces of the Au nanoparticles evaporate or evolve faster than others to cause these discontinuous shrinkage patterns.  Our KMC results give us an atomistic picture of how these facet relationships form and evolve using the varied coordination number of surface atoms as a proxy for anisotropic surface energy. The simulations, in agreement with the mechanism proposed above, show that the desorption process in which a surface atom is removed to form an adatom on the (111) terrace (leading to evaporation or surface diffusion) is most likely initiated from the edges and corners between facets since these sites have the lowest coordination number.  After removal from the edge site, adatoms are most likely to diffuse along the surface rather than directly evaporate due to the high activation barrier for evaporation; this leads to the presence of mobile atoms on the crystal surfaces in equilibrium with the facets themselves.  These surface adatoms are even more under-coordinated after leaving the edge sites and migrate in a random walk until they either find a low energy binding site (near a surface defect, forming a cluster with other adatoms, or at step edges) or evaporate\cite{pimpinelli_what_1994}.

\subsection{Nanoparticle Surface Roughening}
Based on our results, which show experimentally that evaporation begins around 750~$^\circ$C and, computationally that the evaporation process depends on a morphological departure from a smooth surface structure, we expect that the ability of Au atoms to evaporate is closely related to the roughening transition.  While the actual appearance of a discontinuity in surface roughness to signify the roughening transition is debated in the literature, our combination of experimental and simulated data show the presence of rough crystal surfaces and rapid fluctuations in surface structure that cause roughening\cite{zhdanov_surface_1998}. Moreover, our measurements of evaporation rate as a function of temperature from experimental data appear nearly discontinuous, with a clear onset of evaporation near 750~$^\circ$C.  DFT confirms that the energetic barrier for adatom creation is low for the high-index surfaces which form through the evolution process.  This reinforces our understanding that rough surfaces, which form in part through the creation of \{1 1 1\} adatoms from surface edges, are critical to the evaporation process throughout the temperature range we considered.  At higher temperatures the equilibrium will shift such that adatoms on the (111) surface are actually more stable than the low-coordinated surfaces themselves, eventually enabling rapid morphology change and evaporation. With this in mind, though our KMC models assumed a perfect FCC lattice and initial morpholgy determined byt the equiilibrium Wulff construction, we suspect that the evolution behavior is not strongly dependent on the initial configuration of the nanoparticle.  In all cases, we observe surface roughening and that evolution is driven by evaporation of adatoms on rough surface, not by desportion from specific facets, edges, or corners.
Researchers have previously used \textit{ab initio} modeling to map the morphological phase diagram of Au nanoparticles as a function of size\cite{barnard_nanogold_2009}.  They showed, that for particle sizes seen in our experiments, then onset of roughening occurs around 600~$^\circ$C and that by 850~$^\circ$C the nanoparticle should be “quasi-molten”.  Based on this, our experiments essentially probe the entire temperature range over which rough surfaces are expected for small gold nanoparticles.  Our experimental data do not show evidence for immediate and spontaneous roughening or surface melting at any temperature, however, we do see rapid changes in the surface structure which signal the progression between the perfect faceted surface and the high-temperature rough configuration.  

Roughening also decreases the mean coordination number of surface atoms compared to the clean crystal surface.  Considering the physics behind our KMC model, where the probability of desorption/evaporation is inversely related to the coordination number, we can see how the roughest surfaces at the highest temperatures are exponentially more prone to evaporation due to thermal activation and large concentrations of under-coordinated adatoms.  Previous KMC studies have shown the link between evaporation and surface roughening, and described the acceleration of roughening and evaporation in systems where mass is not conserved\cite{zhdanov_surface_1998}. In a different system, it has been shown that the presence of O$_2$ gas reduces the difference in surface energy between varied Pt facets, and leads to the formation of round particles with rough surfaces\cite{altantzis_three-dimensional_2019}. Similarly, research has suggested that the desorption and surface diffusion PtO$_x$ species from rough nanoparticle surfaces are responsible for rapid coarsening \textit{via} Ostwald Ripening for alumina-supported Pt nanoparticles\cite{simonsen_direct_2010}.  As gold is more noble than Pt, even at the nanoscale, the rough nanoparticle surface is subjected to a driving force for evaporation which comes from the pressure difference between the ambient vacuum and the high Gibbs-Thomson pressure at the particle surface, perhaps explaining why we see evaporation rather than any particle growth in our experiments.

\section{Conclusions}
Our use of a combination of experimental and computational techniques has allowed us to understand the mechanisms and rates of supported nanoparticle evaporation as a function of temperature and particle surface characteristics.  Using time-resolved \textit{in situ} TEM, we were able to observe the real behavior of systems containing hundreds of supported gold nanoparticles at various processing temperatures, and track their evolution in time.  Motivated by experimental results which signaled that an activated mechanism drives nanoparticle evaporation, KMC simulations allowed us to suggest atomistic pathways that enable evaporation alongside morphological changes observed experimentally.  Finally, DFT calculations were used to reinforce the link between surface morphology and propensity for degradation.  These calculations show that evaporation occurs through a two-step mechanism where, first, high-energy edge atoms form adatoms on larger \{1 1 1\} terraces, and subsequently the high temperature in the system enables evaporation with an activation barrier confirmed in our experimental analysis.  While our experimental and computational results focus specifically on the evolution of Au nanoparticles, our approach - utilizing common experimental techniques and simulations which include minimal material-specific assumptions - is generalizable to other systems.

With our model of the process by which high system temperatures induce surface roughening and subsequent evaporation, and the knowledge that this process initiates at the edges of high-index surfaces, we can move towards understanding subtleties of supported catalyst degradation without relying on simplifications and assumptions about nanoparticle shape and surroundings.  In this way, findings like ours, which suggest important parameters for catalytic particles, can inform research in materials synthesis to create novel catalyst materials which are less susceptible to degradation under harsh reaction conditions.

\section{Methods}
\subsection{\textit{In situ} TEM Experiments}
Au nanoparticles coated with dendrimer ligands to enable self-assembly at specified particle spacings were synthesized according to previous study\cite{elbert_design_2017}. The nanoparticles used in this experiment had a monodisperse particle size distribution centered at R = $2.1 \pm 0.3$ nm.  After synthesis, nanoparticles were drop-cast directly on to Hummingbird Scientific \textit{in situ} heating chips\cite{noauthor_mems_nodate}.  After drop-casting, heating chips were plasma cleaned under O$_2$ (forward target power = 25~W) for 15 seconds to remove the ligand coating before heating.  Finally, the chip was loaded into a Hummingbird Scientific heating holder and inserted directly into a JEOL F200 S/TEM operating with accelerating voltage of 200~kV.  Chip windows were scanned to find regions containing large areas of self-assembled nanoparticles, and the samples were subsequently heated to temperature.  Images were collected every five seconds for at least one hour using a Gatan OneView electron detector with 4k pixel resolution.  Experimental temperatures range from 650 – 850~$^\circ$C in 50~$^\circ$C increments.

\subsection{Image Segmentation and Analysis}
After each experiment, an unsupervised CNN pipeline was used to segment images and extract particle sizes and positions as a function of time; background information, machine learening Python code, and an interactive image segmentation tutorial are provided in the references\cite{horwath_understanding_2020,vyas_tutorial_nodate}.  A small set of experimental images (~5 images) were randomly chosen from each dataset to build a training set for an unsupervised autoencoder model and each 4k image was cropped into non-overlapping 128 x 128-pixel sections.  Each subsection was then rotated by 90°, 180°, and 270° to augment the dataset .  Since CNN are translation-invariant and recognize image features regardless of their position, each of these image subsections and rotations represents a unique training example.  The unique UNet-type unsupervised CNN was retrained for each experimental dataset, and then applied to raw images using the PyTorch framework\cite{paszke_automatic_nodate}.  Particle sizes and positions were extracted from binary images using the scikit-image connected components algorithm\cite{van_der_walt_scikit-image_2014}.  After processing all experimental datasets, a comprehensive Pandas DataFrame was used to store all experimental data as a function of particle size, time, and temperature.  All further analysis and visualization were performed in Python, and made use of numpy, matplotlib, and scipy libraries\cite{harris_array_2020,hunter_matplotlib_2007,virtanen_scipy_2020}.  Jupyter notebooks demonstrating the analysis and production of figures are available on github at: \url{https://github.com/jhorwath/FacetEvaporation.git}.

\subsection{Kinetic Monte Carlo Simulations}
A homemade script was used for kinetic Monte Carlo (KMC) simulations (simulation scripts are available on the github repository listed above).  To begin, a NP  model was made by sampling the experimental particle size distribution, and forming the equilibrium FCC structure for the particle based on the Wulff construction applied in the Atomic Simulation Environment (ASE) python package to use as simulation input\cite{Larsen2017}.  KMC event probabilities were calculated according to Equation \ref{eq:event_probability}: 
\begin{equation}
    P_{event} = v_0 \exp\left[\frac{-\epsilon N}{k_B T}\right]
    \label{eq:event_probability}
\end{equation}
where $v_0$ is an attempt frequency chosen to match the experimental physics, $\epsilon$ is an energy per bond (taken as 0.1 eV / bond) multiplied by coordination number $N$\cite{pyykko_theoretical_2004}. $k_B$ and $T$ are Boltzmann's constant and temperature, respectively.

In each KMC step, the event probability was calculated for each atom in the particle to determine which atoms will move/evaporate\cite{he_lattice_2016,sickafus_introduction_2007,jansen_introduction_2012,landau_guide_nodate}.  In order to accelerate the simulation, only surface atoms ($N < 12$) were considered in KMC time steps.  For each event, all available lattice sites surrounding an atom were considered as possible diffusion sites, and evaporation was considered as diffusion into the vacuum.  To determine which move to carry out, a rate catalog based on the activation barrier for each candidate was built, and then sampled using a random number generator.  The diffusion event rate, $r$, was  determined by Equation \ref{eq:rates}:
\begin{equation}
    r = v_0 \exp \left[\frac{-\Delta E}{k_B T}\right]
    \label{eq:rates}
\end{equation}
where $\Delta E_{diffusion} = -\epsilon (N_1 - N_0)$ for coordination numbers $N_1$ of the candidate site and $N_0$ at the current site.  Correspondingly, $\Delta E_{evaporation} = \epsilon N_0$ since evaporation requires breaking all surface bonds.  By this model, diffusion to a more coordinated lattice site is favorable, and evaporation occurs with low probability but is still possible for atoms with low coordination number at high temperatures.  Simulations were run until all atoms in the system evaporated.  To ensure that KMC results matched the expected physical behavior, KMC evaporation rates were plotted to show that $\frac{dR}{dt} \propto \frac{1}{R}$, in agreement with our previous research on nanoparticle evaporation (Supplemental Figure 4)\cite{horwath_quantifying_2021}.  
3D visualizations of KMC trajectories were made using the Plotly python library\cite{inc_collaborative_2015}.  Diffraction simulations of KMC nanoparticles were carried out using the Debye scattering equation and were performed using the XrDebye code available in the ASE python package.  

\subsection{Density Functional Theory}
For the DFT work, all systems are built using the Atomic Simulation Environment (ASE) \cite{larsen_atomic_2017} and all calculations are performed using Quantum Espresso's \cite{giannozzi_quantum_2009} implementation of generalized gradient approximation (GGA) DFT using the revised Perdew-Burke-Ernzerhof (RPBE) exchange-correlation functional \cite{Hammer1999} and a plane wave cutoff of 400 eV. A 5x5x1 Monkhorst-Pack grid \cite{Pack1977} is used for k-point sampling on all relaxations. Periodic boundary conditions are implemented with a dipole correction \cite{Bengtsson1999}. All calculations are converged to 10\textsuperscript{-5} eV and forces are converged to 0.03 eV/\AA. Core electrons are treated using ultrasoft pseudopotentials \cite{Vanderbilt1990a}. Fermi-Dirac smearing is used with a smearing width of 0.1 eV. Calculations are not spin-polarized.

Entropy corrections are calculated using the ASE Thermochemistry package \cite{Larsen2017}. Adatoms on the (111) surface are treated using the Hindered Translator method \cite{Sprowl2016} with a diffusion barrier of 0.01 eV (equal to the difference in adsorption energies for a gold adatom on the threefold and bridge sites of the (111) surface) and only translational entropy is included. Surface lattice atoms are treated as harmonic oscillators with only vibrational entropy contributions. Gas phase atoms are treated as ideal gas with an entropy of 0.002 eV/K. 

DFT calculated ground state energies are then corrected with these entropies to get the free energy of the system according to Equation \ref{eq:free_en}:
\begin{equation}
    G = E_{DFT} - TS
    \label{eq:free_en}
\end{equation}

\section{Acknowledgments}
J.P.H. and E.A.S acknowledge support through the National Science Foundation, Division of Materials Research, Metals and Metallic Nanostructures Program under Grant 1809398.  We thank Katherine Elbert, Shengsong Yang, and Christopher Murray from the University of Pennsylvania Department of Chemistry for help with nanoparticle and dendrimer ligand synthesis.  Additionally, we acknowledge the support of Peter Voorhees, of the Northwestern University Department of Materials Science and Engineering, for guidance while developing KMC models and discussion of surface evolution kinetics. C.L.C. would like to thank the Vagelos Institute for Energy Science and Technology (University of Pennsylvania) for support through a graduate fellowship. C.L.C. and A.V. acknowledge the use of the computer time allocation at the Extreme Science and Engineering Discovery Environment (XSEDE) supported through National Science Foundation Energy under Award Number DMR180108.

\newpage
\bibliography{Jay_Zotero.bib,Gold_NPs-methods.bib}

\end{document}